\documentclass[a4paper,fleqn,usenatbib]{mnras}

\usepackage{newtxtext,newtxmath}
\usepackage[T1]{fontenc}
\usepackage{ae,aecompl}

\usepackage{graphicx}	
\usepackage{amsmath}	
\usepackage{amssymb}	
\usepackage{color}      
\usepackage{subcaption}  
\usepackage{hyperref}   
\usepackage{verbatim}
\usepackage{float}

\def\lsim{\mathrel{\lower0.6ex\hbox{$\buildrel {\textstyle <}
 \over {\scriptstyle \sim}$}}}
\def\gsim{\mathrel{\lower0.6ex\hbox{$\buildrel {\textstyle >}
 \over {\scriptstyle \sim}$}}}

\begin{document}
\label{firstpage}
\title[The Origin of Lopsided Satellite Galaxy Distribution in Galaxy Pairs]{The Origin of Lopsided Satellite Galaxy Distribution in Galaxy Pairs}

\author[Chen Chris Gong et al.]
{Chen Chris Gong$^{1,2}$\thanks{E-mail: cgong@uni-potsdam.de}, Noam I Libeskind$^{2,3}$,
Elmo Tempel$^{4,2}$,Quan Guo$^{5,2}$, \newauthor
Stefan Gottl\"{o}ber$^2$, 
Gustavo Yepes$^{6}$,
Peng Wang$^{2}$,
Jenny Sorce$^{7,2}$, Marcel Pawlowski$^{2}$\\
$^{1}$University of Potsdam, Institute of Physics and Astronomy, Karl-Liebknecht-Stra\ss{}e 32, 14476 Potsdam\\
$^{2}$Leibniz-Institut f\"ur Astrophysik Potsdam (AIP), An der Sternwarte 16, D-14482, Potsdam, Germany\\
$^{3}$University of Lyon; UCB Lyon 1/CNRS/IN2P3; IPN Lyon, France\\
$^{4}$Tartu Observatory, University of Tartu, Observatooriumi 1, 61602 T$\tilde{o}$ravere, Estonia\\
$^{5}$Shanghai Astronomical Observatory (SHAO), Nandan Road 80, Shanghai 200030, China\\
$^6$ Departamento de F\'isica Te\'{o}rica and  Centro de Investigaci\'{o}n Avanzada en F\'isica Fundamental (CIAFF), Facultad de Ciencias,\\
~ Universidad Aut\'{o}noma de Madrid, 28049 Madrid, Spain\\
$^7$ University of Lyon, University of Lyon 1, Ens de Lyon, CNRS, Centre de Recherche Astrophysique de Lyon UMR5574, F-69230, Saint-Genis-Laval, France
}

\pubyear{2018}

\maketitle

\begin{abstract}
It is well known that satellite galaxies are not isotropically distributed among their host galaxies as suggested by most interpretations of the $\Lambda$CDM model. One type of anisotropy recently detected in the SDSS (and seen when examining the distribution of satellites in the Local Group and in the Centaurus group) is a tendency to be so-called ``lopsided''. Namely, in pairs of galaxies (like Andromeda and the Milky Way) the satellites are more likely to inhabit the region in between the pair, rather than on opposing sides. Although recent studies found a similar set up when comparing pairs of galaxies in $\Lambda$CDM simulations indicating that such a set up is not inconsistent with $\Lambda$CDM, the origin has yet to be explained. Here we examine the origin of such lopsided setups by first identifying such distributions in pairs of galaxies in numerical cosmological simulations, and then tracking back the orbital trajectories of satellites (which at $z=0$ display the effect). We report two main results: first, the lopsided distribution was stronger in the past and weakens towards $z=0$. Second, the weakening of the signal is due to the interaction of satellite galaxies with the pair. Finally, we show that the $z=0$ signal is driven primarily by satellites that are on first approach, who have yet to experience a ``flyby''. This suggests that the signal seen in the observations is also dominated by dynamically young accretion events. 
\end{abstract}


\section{Introduction} \label{sec: intro}

The favoured scenario for how structures in the Universe form is known as the $\Lambda$CDM paradigm, wherein the universe is composed primarily of dark matter ($\sim26\%$), dark energy ($\sim70\%$) and baryons ($\sim4\%$) \citep{Planck13}. Accordingly, small perturbations in the otherwise nearly homogeneous density field, decouple at high red-shift from the expansion of the Universe and collapse to form the first dark matter haloes. Dark matter haloes proceed to grow hierarchically via gravitational instability \citep{Zeldovich80} in a ``bottom-up'' fashion: small haloes merge to form larger ones and so on \citep{WhiteRees78}. 

One of the defining features of this scenario is that the dark matter haloes are continually merging with one another. Because of the nature of the power spectrum of fluctuations measured in the CMB, the majority of the merger events will involve small, low mass haloes \citep{LaceyCole99}. Indeed studies such as \cite{Genel10} have shown that Milky Way mass haloes gain just $20-30$\% of their mass via discrete resolved mergers (with a mass ratio of less than 10:1) and 70-80\% via smooth ambient accretion \citep[but see also:][]{FakhouriMa08,Stewart08,Madau08,AnguloWhite10,LHuillier12}. Often the dynamical time needed for such small haloes to lose enough angular momentum so that tidal forces can rip them apart them is quite long \citep[e.g. equation 2 in][]{WetzelWhite10}. As a result small ``satellite'' galaxies exist in abundance in the nearby environment of larger galaxies \citep{Tollerud08,koposov2008,Newton2018}.

Such dwarf galaxies are observed to cluster in the vicinity of larger galaxies \cite[]{Pawlowski12, Mu_oz_2015, Ordenes_Brice_o_2018, Taylor_2018}. The geometry of this clustering has been a subject of investigation at least since \cite{Holmberg1969} identified an anisotropy in the distribution of satellites surrounding disk galaxies. Indeed, there is an open debate in the literature on if satellite galaxies are distributed isotropically based on observation \citep[e.g.][among others]{Kroupa05,2013Natur.493...62I,RIbata14,Pawlowski14,Pawlowski2018rev, Tully15,Cautun15,Muller16,Bullock17,Arakelyan18}; and if the satellite galaxies are distributed anisotropically, whether or not there is a persistent sense of orientation or rotation among them \citep{MetzKroupaLibeskind2008,Pawlowski13,NIbata14,Mueller18}; and when there is, whether such a structure could maintain itself (\citealt{Bowden2014}). Most of these studies have explicitly focused on the anisotropy of satellite distributions of field galaxies (as opposed to galaxies in clusters or pair of galaxies). Yet recently, the anisotropic distribution of the satellites of galaxies pairs has been brought to attention, which has been seen in the Local Group as well as in the Centaurus group (\cite{Mueller15}, \cite{Mueller17}).

The existence of galaxy pairs (such as the two main galaxies of the Local Group and those of the Centaurus group) presents the cosmologist with a unique testing ground for questions related to structure formation and the cosmological model. This has been recognised at least since the timing argument of \cite{KahnAndWoltjer59} who famously used the dynamics of the Local Group to compute its mass according to a cosmological paradigm \citep[see also extensions such as][]{PLH13}. Indeed in recent years there has been an abundance of work that examines in detail the properties of pairs in both large sky surveys \citep[e.g.][to name a few]{2008AJ....135.1877E,2016RAA....16...43S,2015MNRAS.450.1546B,2015Tempel,2017MNRAS.465.2671G} or in numerical simulations \citep[e.g.][and references therein]{2013MNRAS.436.1765M,2009MNRAS.397..748P, Garrison-Kimmel}. The vast majority of these studies have focused on pairs for the purpose of understanding the physics of galaxy formation and galaxy conformity \citep{2013MNRAS.430.1447K}, namely how living in a pair (and the eventual merger) affects a galaxy's star formation efficiency, gas reservoir and morphology. 

A number of recent projects have extended the research on satellite galaxy distribution from those around isolated field galaxies towards those which inhabit galaxy pairs, and have compared the distribution of small satellites in these two types of environment when their hosts are of similar mass and other characteristics. Since only a handful of such pairs have been both spectroscopically confirmed and imaged with sensitivity down to the smallest dwarfs (for example the Centaurus A - M83 pair), most people wishing to examine satellites of galaxy pairs must turn to large photometric catalogues \citep[e.g.][among many others]{2019MNRAS.484.4325W} and employ stacking. 

\cite{Libeskind16} examined the distribution of satellite galaxies near pairs of galaxies that resemble the ``Local Group'' (in a very broad sense) identified in the SDSS. Since each pair in such a survey has just a few satellites, any trend in the distribution of satellites is revealed by stacking tens of thousands of such pairs. They recovered the well known result that satellites are distributed anisotropically or ellipsoidally (and not spherically) among their hosts \citep[see also][]{2004MNRAS.348.1236S, Libeskind05, 2010ApJ...709.1321A, TempelKipper15, Libeskind15, 2015MNRAS.449.2576C}. The analysis further revealed that satellites of galaxy pairs identified in the SDSS show a statistically significant tendency to bulge towards the partner galaxy, in a lopsided manner. This effect was coined ``the Lopsided distribution of satellite galaxies'' owing to its ovular or egg-like configuration. While the small satellites are found to preferentially inhabit the region between the pair, this striking effect was shown not to be a result of projection effects or simply the overlapping of two independent satellite distributions. It should be noted that the satellites of Andromeda display exactly the same lopsidedness with the majority on the near side of M31. (\cite{Bowden2014} addressed this issue analytically ruling out tides as being responsible, but suggesting that recent group accretion may explain the effect.) Interestingly after the publication of \cite{Libeskind16}, \cite{2017MNRAS.468.2605E} found a very similar signal by stacking lensing images of pairs, albeit on slightly larger scales \citep[see also][]{deGraaff17}.

Since then, \cite{Pawlowski17} has found such a lopsided effect of satellite galaxy distribution in different $\Lambda$CDM simulations confirming that such an arrangement is a feature of the cosmological paradigm, possibly even a generic one. Although the strength of the signal found in simulations is not identical to that found in observations, the discrepancy can easily be attributed to the slight differences in the samples being compared, and to fore- and background contaminations in the observed sample. \cite{Pawlowski17} noted that although successful in finding such a lopsided distribution, the evolutionary history and causal mechanism for the lopsidedness was left unexplained. 

To some degree, in this study, we pick up where \cite{Pawlowski17} left off. We conduct an investigation into the origin of the lopsided satellite distribution in galaxy pairs like the Local Group which are found in the cosmological simulations. Due to the nature of our study we are then able to track these lopsided distributions back in time and investigate their origin. In doing so we characterize the accretion history of the satellites and their host haloes. 

\section{Methods} \label{sec:methods}
In this section we describe how our sample of halo pairs (and their satellites) are identified.

\subsection{Simulation and the identification of haloes and subhaloes} \label{sec: Rockstr}
For our analysis, we use the ``Extremely Small Multidark'' simulation (ESMD). ESMD is an extension of the Multidark suite of pure $N$-body simulations\footnote{ see https://www.cosmosim.org for more detailed information.} \citep{Klypin16}.
This is a dark-matter-only $N$-body simulation assuming a $\Lambda$CDM power spectrum of fluctuations according to the Planck cosmological  parameters (\citealt{Simulation};  $\Omega_{\Lambda} = 0.69$, $\Omega_{M} = 0.33$, $\sigma_{8} = 0.83$ and $H_{0} = 67.77$ km s$^{-1}$ Mpc$^{-1}$). The simulation consists of $N_{\rm part}=2048^{3}$ particles in a periodic box of side length $L_{\rm box} = 64$ $h^{-1}$Mpc. Such a simulation achieves a mass resolution of $2.6\times10^{6}~ h^{-1}$M$_{\odot}$ per particle and a spatial softening length of $1h^{-1}$kpc.  

The haloes and subhaloes in the simulation are identified using the publicly available  \href{https://bitbucket.org/gfcstanford/rockstar}{\sc Rockstar} halo finder (Robust Over-density Calculation using K-Space Topologically Adaptive Refinement) which is described in detail in \cite{Rockstarhalofinder}. We briefly summarise how {\sc Rockstar} works but refer the reader to \cite{Rockstarhalofinder} for details.  {\sc Rockstar} is a massively parallel method of identifying haloes and their substructures based on adaptive hierarchical refinement of friends-of-friends (FOF) groups in six phase-space dimensions and one time dimension.

The halo finder begins by identifying 3DFOF groups based only on positional information. Then with these 3DFOF groups, {\sc Rockstar} adaptively chooses a phase-space linking length based on the standard deviations of the particle distribution in position and velocity space. In this way subhaloes can be identified and are then assigned to their hosts similar to the way particles are assigned to their 6DFOF subgroups. Additionally, particle-based merger trees are created. For a given halo, its descendant is assigned as the halo in the next time step which share the maximum number of common particles. We consider only haloes resolved with more than 20 particles, resulting in a minimum halo mass of $M>5.2 \times 10^{7} h^{-1}M_{\odot}$. Although uncertain due to the stochasticity of star formation processes at these low masses, such dwarfs could host galaxies with a stellar mass below $10^{4}$ (\citealt[]{Sawala14, Reed17}) or could have star formation entirely suppressed (\citealt{dwarfmass2017}).

\begin{figure}
\captionsetup[subfigure]{justification=centering}
\centering
 \includegraphics[width=.3\textwidth]{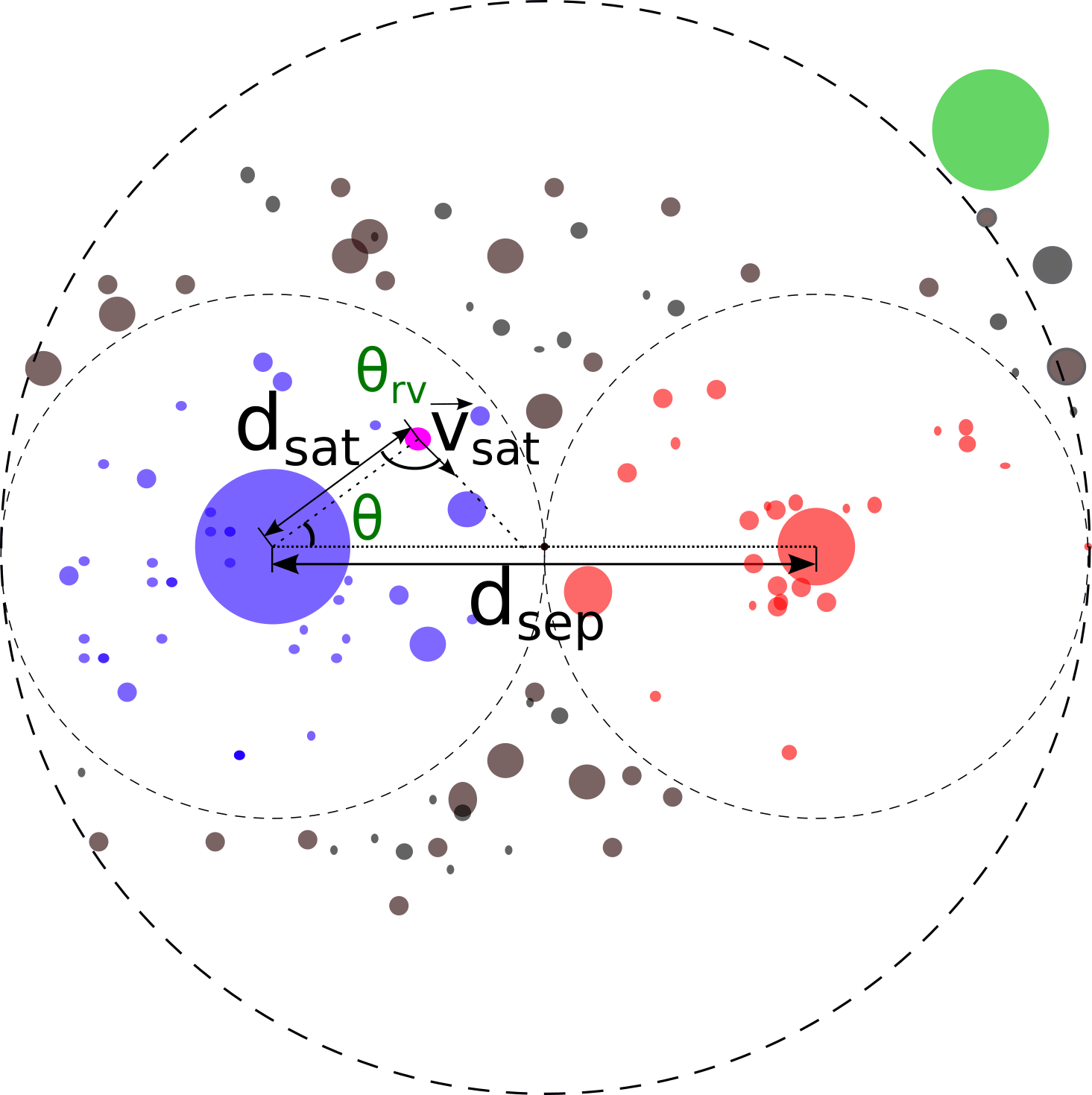}
 \caption{A sketch of a typical binary halo system. The pair is separated by a distance $d_{\text{sep}}$. Haloes or subhaloes that are within 1/2 the separation distance, i.e. $d_{\rm sat} < 0.5 d_{\text{sep}}$ from a host are identified and referred to as satellites of that host (colored red and blue here). For illustration, the position angle $\theta$ of a satellite (in magenta) of its host (in blue) is measured between the line connecting the pair and a satellite's position vector. $\theta_{\text{rv}}$ indicates the angle between the velocity of the satellite ($\vec{v}_{\rm sat}$) and its position angle and is a measure of how radial the infall is. The bigger circle indicates the exclusion range (see criteria \ref{3rdcriteria}), beyond which a third halo which is more massive than the less massive member (red) of the pair cannot trespass (e.g. the halo in green). \label{fig:bindraw}} 
\end{figure}

\subsection{Identification of halo pairs and satellites} \label{sec: haloid}
Loosely based on the mass and geometrical set up of the Local Group, we identify halo pairs at  $z=0$ using the following criteria: 
\begin{enumerate}
\item halo mass range: $10^{11}<M/(h^{-1}M_{\odot})<5\times10^{12}$ 
\item halo separation:  $0.3 < d_{\text{sep}}/ (h^{-1} {\rm Mpc}) < 1.5$
\item Exclusion: There is no third halo more massive than 1/2 of the mass of the less massive host halo within a radius, centered on the binary's midpoint, and equal to the separation distance from the binary center. \label{3rdcriteria}
\end{enumerate}

We note that the values for the halo mass, halo pair separation and exclusion mass are allowed to be within chosen ranges, which are loosely motivated by the values seen in the Local Group. They are deliberately chosen, where we try to both maintain a physical justification (approximating, even loosely, the situation in the Local Group) and to obtain a maximal sample size. At $z=0$, our criteria returns $2~252$ halo pairs with  $625~033$ satellites among them. The geometry of satellites in such a halo pair is shown in Fig.~\ref{fig:bindraw}. 

For each halo pair, the smaller haloes and subhaloes around them are identified as satellite haloes within $0.5 d_{\text{sep}}$ from its nearest host halo. These include both subhaloes located within their host's virial radius and ``field dwarfs" located outside the virial radius; we collectively term all of these ``satellite haloes''. Satellite position may be characterised by the angle, $\theta$, made between the satellites position vector from the closest primary and the line connecting the binary. The viewing angle is rotated such that $\theta$ is maximum, and the satellite -- host -- partner triangle is viewed ``face on''. We treat all satellites equally regardless of mass. In effect our analysis proceeds by considering the dynamics of each satellite halo as it is accreted by its host halo pair. 

\subsection{Control (Overlapping) sample at $z=0$ constructed from two isolated haloes}\label{subsec: control}
Following similar reasoning as in \cite{Libeskind16}, when the distribution of $\theta$ is examined, any non-uniformity must be separated from the possible effect of overlap. Namely, two haloes with spherically isotropic satellite distributions, when brought near to each other, would naturally show an increased number density of satellites in regions where their satellite distributions overlap, which happens purely geometrically, without the influence from the force of gravity. This follows works such as \cite{Prada_2006} who have shown that haloes of the mass range we are considering are extended up to $2\sim 3$ virial radius.

In order to quantify the effect of overlap, halo pairs are artificially constructed from isolated haloes and their satellites. To do this, we must match each member of a pair to an isolated halo. The two isolated haloes and their satellites are then artificially brought together and placed at the same separation distance as the real halo pair. In detail, for each halo pair, we note the pair separation $d_{\rm sep}$. We then identify two haloes in the same mass range (i.e. $10^{11} - 5\times10^{12}h^{-1}M_{\odot}$) that are isolated and are the closest in mass to the members of the real halo pair, respectively. Here, ``isolated'' means that the nearest halo, which is either more massive than the less massive member of the pair, or more massive than the ``isolated'' halo itself, is at least $1.5\times d_{\rm sep}$ away. This ensures that when we bring the two haloes together and place them at distance $d_{\rm sep}$ apart, there is no third halo bigger than either member of the real halo pair and no third halo bigger than either member of the artificial halo pair. 
Once two isolated haloes are matched to a real pair, their satellites are chosen in the same way as in the main sample, namely all haloes within $d_{\rm sep}/2$ of each halo's center are identified. The isolated haloes (and their satellites) are then placed at a distance of $d_{\rm sep}$ from each other, therefore creating an artificial pair composed of two isolated haloes. 

\begin{figure}
\centering
\includegraphics[width=.48\textwidth]{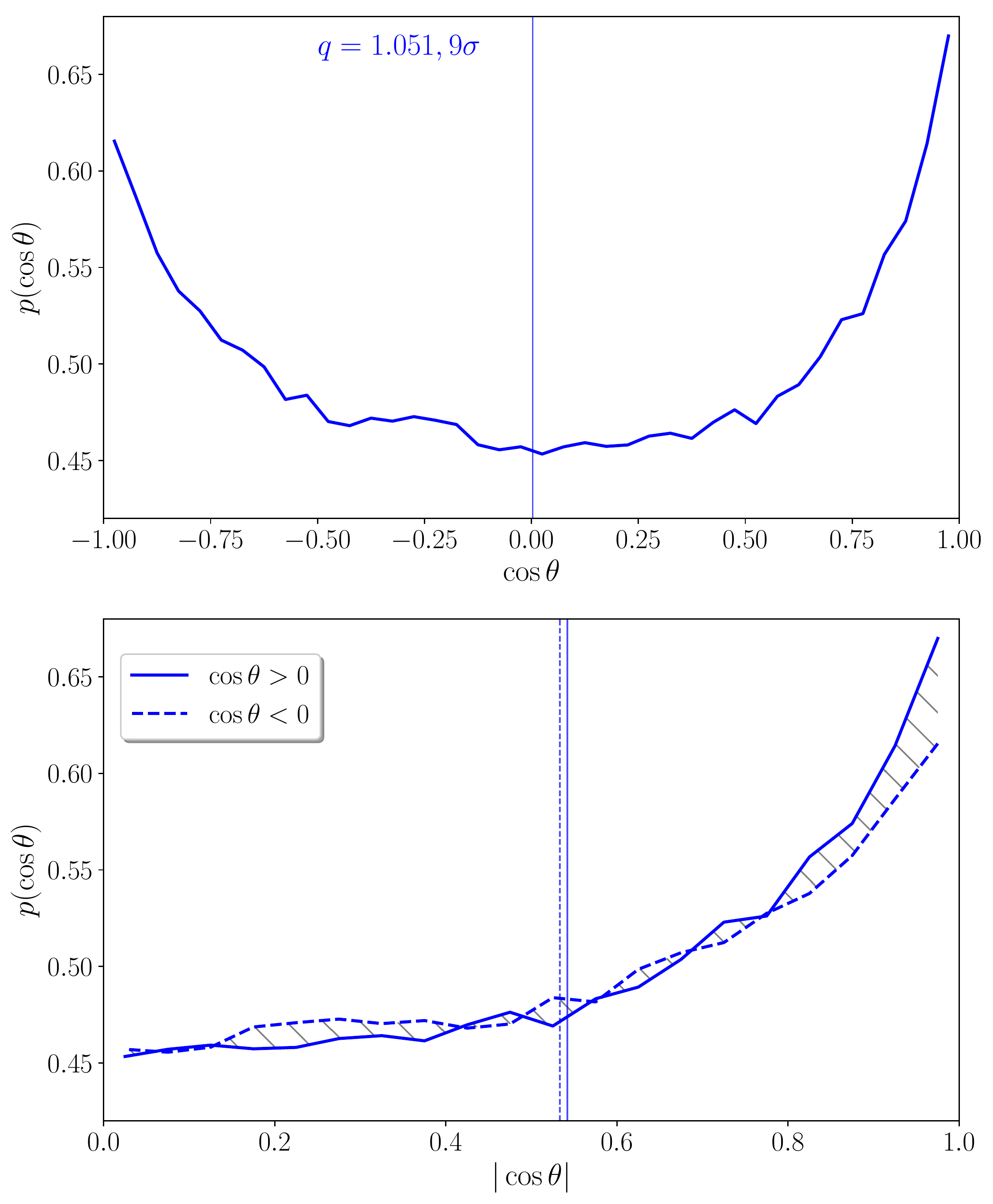}\\
\caption{The lopsided signal is characterised by the probability distribution of $\cos\theta$ shown in top panel. In bottom panel, in order to emphasise the lopsided signal we divide the distribution into two regions: the ``facing'' (i.e. $\cos\theta>0$, solid) and ``facing away'' region from the other primary halo (i.e. $\cos\theta<0$, dashed), and plot the two distributions of absolute value of $\cos\theta$ for direct comparison. It is obvious that the solid line dominates for $|\cos\theta| >0.8$, indicating there are more satellites close to the line connecting the pair in the region in between the pair than 180 degrees away. The thin vertical lines denote the medians of the distributions which are although close, are statistically inconsistent with a symmetric (non-lopsided) distribution. The lopsidedness metric $q$ and its statistical significance $\sigma$ are defined at the end of section \ref{sec: signal}. In short, $q$ is the ratio between the number of satellites in the region with $\cos > 0.8$ and the number of those in $\cos<0.8$. Its $\sigma$ is computed from the variance in this quantity from random trials. \label{fig:binwsigma}} 
\end{figure}

\section{Results} \label{sec:result1}
\subsection{Lopsided $z=0$ signal} \label{sec: signal}

We begin by re-examining the findings by \cite{Pawlowski17}, who confirm the observational result of \cite{Libeskind16} with similar numerical simulations to those used here. In Fig.~\ref{fig:binwsigma} we show the normalized probability distribution of $\cos\theta$ for all satellites of all host pairs in our sample. The peaks of the distribution at $|\cos\theta|\approx 1$ in Fig.~\ref{fig:binwsigma} top panel indicates that the satellites have a strong tendency to be aligned with the line connecting the halo pair in both regions. 

A direct comparison between the two regions ``facing'' the other primary halo and ``facing away'' from the other primary halo show lopsidedness exists for this signal. An excess probability is seen for $|\cos\theta| \gsim 0.8$ for the $\cos\theta>0$ interval with respect to $\cos\theta < 0$. Furthermore the asymmetry can be quantified by the median of the absolute value:
\begin{eqnarray*}
\langle|\cos\theta|\rangle = 0.529 & {\rm for} \cos\theta>0\\
\langle|\cos\theta|\rangle = 0.523 & {\rm for} \cos\theta<0
\end{eqnarray*}
Although this does not seem like a large difference, given our sample size it is strongly statistically significant, as described below.

To consider the statistical significance of the median values stated above, we compare to 10,000 samples of the same size. Namely, we generate $625~033$ random, uniformly distributed numbers in the interval [0,1] (representing a uniform $|\cos\theta|$ distribution). The median of these $625~033$ is then computed. This is then repeated 10,000 times. In this way we compute the distribution (standard deviation) of the median value of $625~033$ random numbers drawn from [0,1] as $\sigma_{\rm m}=0.001.$ Therefore both the facing ($\cos\theta>0$) and facing away ($\cos\theta<0$) samples are inconsistent with a spherical uniform distribution at roughly $\sim 10\sigma_{\rm m}$ level. 

We now use this value of $\sigma_{\rm m}$ to determine how much more asymmetric the ``facing'' signal is than the ``facing away'' signal, determining that the facing sample is more aligned than the facing away sample at roughly the $\sim 2.1\sigma_{\rm m}$ level. This is, in effect, one aspect of the lopsided signal found by \cite{Libeskind16}: satellites facing the opposite halo are more aligned with the line connecting the pair than satellites facing away from the opposite halo. The fact that the satellites in the ``facing region'' are just 2.1$\sigma_{\rm m}$ \cite[and not more as found by][]{Libeskind16}, than their counterparts in the ``facing away'' region, is surprising and demonstrates that both regions include flattened satellite distributions, aligned with the geometry of the pair.

We can also define a lopsidedness metric $q$ that is used in the rest of the paper to quantify the lopsidedness. For a distribution of $\cos\theta$ of size $N$, the number of satellite haloes in the most aligned subsample, namely with $0.8<\cos\theta<1.0$ is termed $N_+$. The number of satellite haloes in the most counter-aligned subsample, namely $-1.0 < \cos\theta < -0.8$ is termed $N_-$. We then define $q = N_+/N_-$: when $q>1$ we have a bulging satellite distribution {\it facing} the other pair member. When $q<1$ we have a bulging distribution {\it facing away} from the other pair member. In order to asses the statistical significance of a given value of $q$, we compute the expected fluctuations in $q$ by generating 10~000 random data sets of the same size, drawn from uniform distribution between $-1$ and $1$,  and getting the expected value of $q$ from these random sets: $\sigma = \operatorname{std}(q)$.

\begin{figure}
\centering
\includegraphics[width=.5\textwidth]{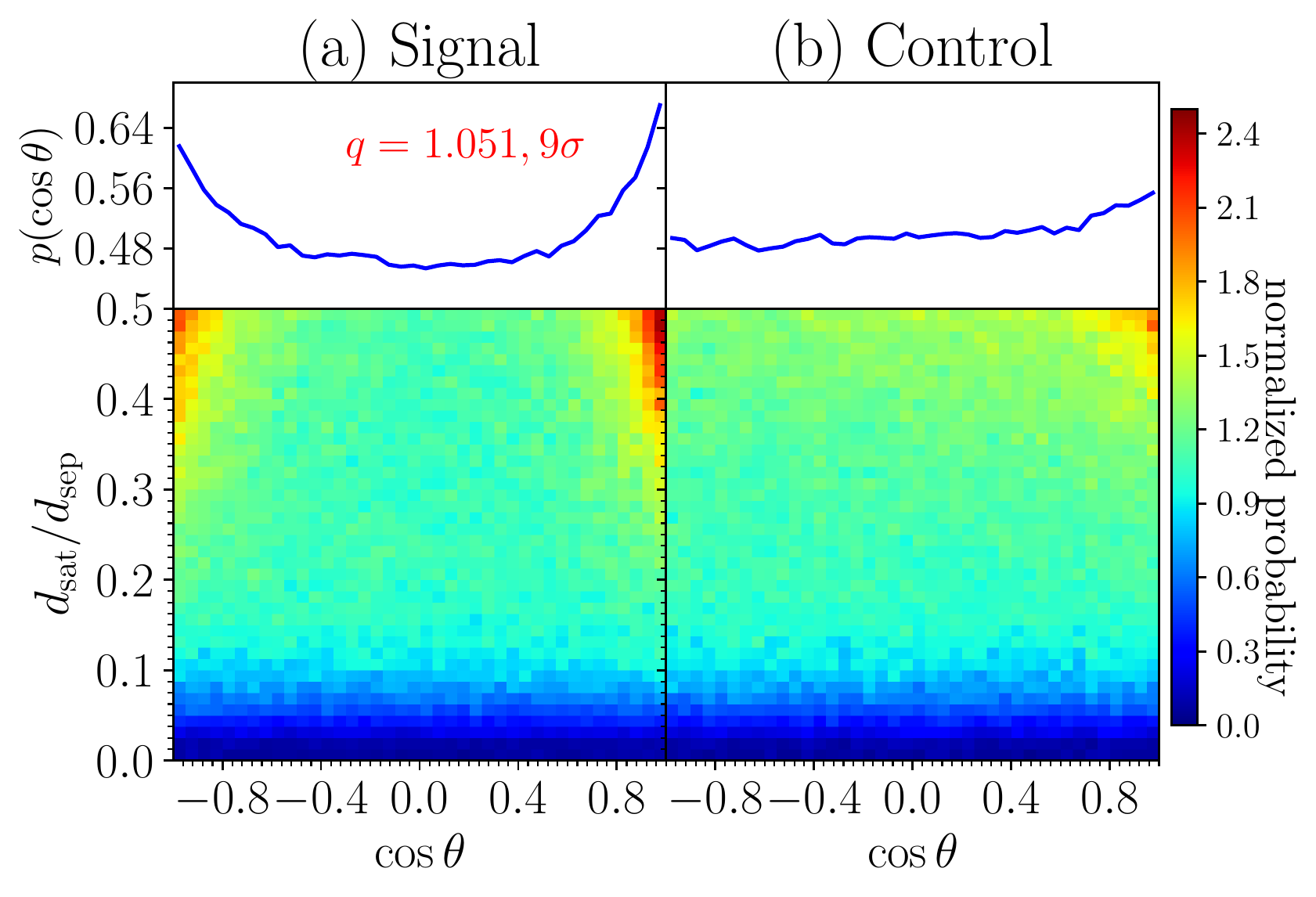}
\caption{We quantify the lopsided signal by showing it as both a probability distribution of $\cos\theta$ (x axis and top histograms) as well as a function of satellite radial position (bottom 2D histogram). Satellite distance to the host is normalized by the pair separation, $d_{\text{sep}}$ (y axis). We show: (a) the signal and (b) control sample of artificial pairs constructed from isolated haloes. \label{fig:2Ddensity_all} } 
\end{figure}

\subsection{Quantifying the lopsided signal at $z=0$.} \label{sec: quantsignal}
\begin{figure}
\centering
    \includegraphics[width=.45\textwidth]{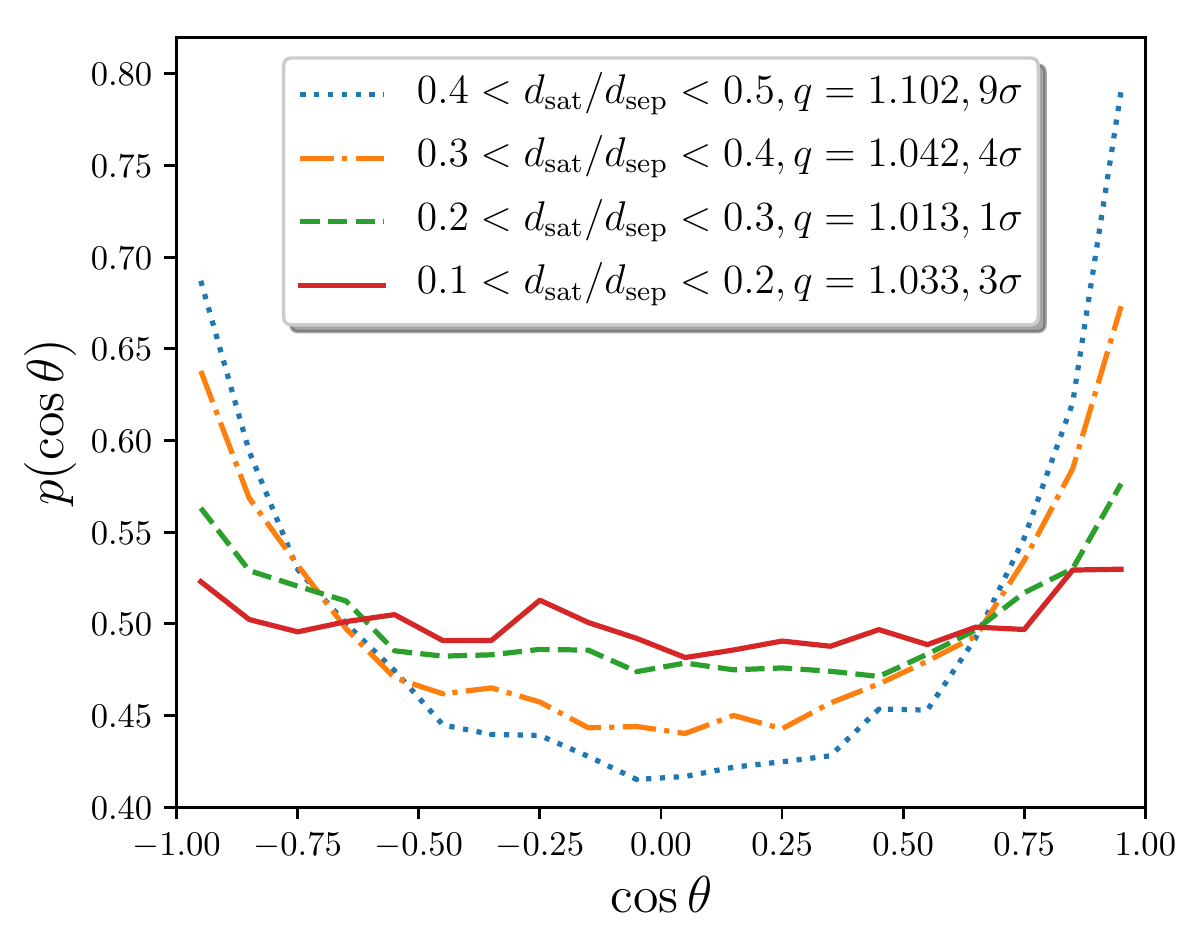}
\caption{Probability distribution of $\cos\theta$ plotted for 4 subsamples of satellites sorted by their distance to hosts. Although the satellite distributions at all distances are anisotropic and flattened, it is only at the greatest distances that they are lopsided.\label{fig:linedensity}} 
\end{figure}

In this section we wish both to study what features of the halo pair might correlate with the lopsided signal and we wish to compare it with the control sample of artificial pairs constructed from isolated haloes (see subsection \ref{subsec: control}). In Fig.~\ref{fig:2Ddensity_all} the cosine angle and radial distribution of satellites are shown in normalized 2D histograms for the sample of pairs (Fig.~\ref{fig:2Ddensity_all}a) and the control sample of artificial pairs (Fig.~\ref{fig:2Ddensity_all}b). Note that the upper panels in this plot show histograms of the probability distribution of the angle (i.e. irrespective of distance of satellite).

Fig.~\ref{fig:2Ddensity_all}a shows that satellites are more likely to be situated along or near the line connecting the binary (namely $\cos\theta \approx\pm1$). Furthermore the lopsided signal is driven by those satellite haloes that are far from their host, near the binary center ($d_{\text{sat}} \approx 0.5 d_{\text{sep}}$, more on this effect in section~\ref{sec:sat_dist}). Note that the  particle resolution inhibits and artificially suppresses the ability for subhaloes to be identified in the densest central regions of a halo (here clearly seen by the regions $d_{\text{sat}} < 0.1 d_{\text{sep}}$). Regardless, Fig.~\ref{fig:2Ddensity_all}a clearly shows that the lopsided signal is driven by the fact that the satellite haloes that are near the binary center also tend to extend along or near the line connecting the two binary members.

Fig.~\ref{fig:2Ddensity_all}b shows the cosine angle and radial distributions of the satellites for the control sample of artificially generated pairs, meant to qualify the effect of overlap. The overlapping effect clearly results in a gradual and slight over-density towards the binary center at $\cos\theta \approx 1$ by construction.  However, the distribution of satellites of the artificially constructed halo pairs is significantly different from the that of the real sample. (A KS test reveals that the two distributions are highly unlikely to be drawn from the same parent sample.) While the real sample is characterised both by a flattening of the satellite distribution aligned with the partner halo as well as by a lopsidedness bulging towards partner, the control sample lacks the flattening but does exhibit lopsidedness due to the overlap effect that we wish to capture. Is this overlap enough to explain the lopsided signal? To some extent it does appear to be of the correct magnitude and the $z=0$ effect could be at least partially ascribed to overlapping satellite distributions. However, such an explanation would suggest that the lopsided signal should decrease with increasing pair separation (i.e. with increasing $z$), which, as we will show later in Section \ref{sec: highz}, is not the case.

\subsubsection{Distance of satellite from host} \label{sec:sat_dist}

In this section we examine how the lopsided signal changes as function of the satellite's distance to the host. We start by examining Fig.~\ref{fig:2Ddensity_all}, where in the top panels we plot the lopsided signal averaged over all distances and in the bottom panels we show a ``heat map'' for the cosine of alignment angle $\theta$ as a function of satellite distance from host. Namely, each satellite provides a $\cos\theta$ and a $d_{\text{sat}}/d_{\text{sep}}$ value, and we bin the satellites in both values. 

Fig.~\ref{fig:2Ddensity_all}a indicates that for the satellites which are more distant from their hosts, their distribution is more anisotropic. This is more explicitly shown in Figure \ref{fig:linedensity}, where we examine the distribution of satellites in concentric shells of increasing distance from their host.

Fig.~\ref{fig:linedensity} shows that the shape of the satellite distribution tends to be oblate yet symmetric (i.e. not lopsided) for the regions closest to the host, and only when the more distant satellites are considered: those at $d_{\rm sat} > 0.3~d_{\rm sep}$, does the distribution start to become more lopsided. Such a finding suggests that it is the opposite binary partner which may be responsible for destroying the aspherical but symmetric satellite distributions, making them more oblate and lopsided. 

\begin{figure}
\centering
{\bf \hspace{.3cm} (a) Less massive hosts \hspace{.5cm} (b) More massive hosts}
\includegraphics[width=.5\textwidth]{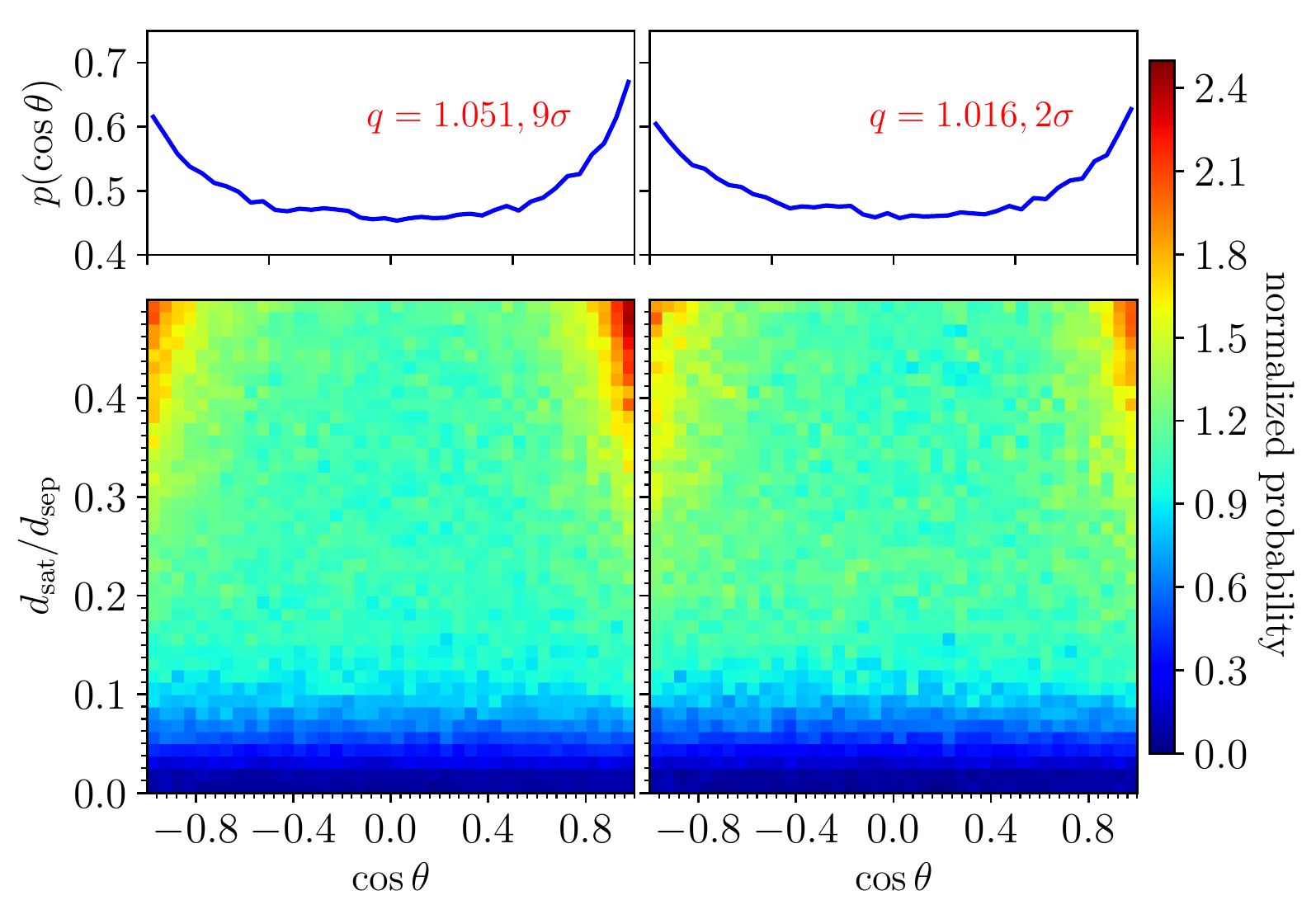}
\caption{We divide the lopsided signal into two subsamples, according to whether the satellite belongs to the less massive member of the pair (panel (a), left) or the more massive one (panel (b), right). All satellites are defined as located within $d_{\rm sep}/2$ from their host. \label{fig:density_bias2}}
\end{figure}

\begin{figure*}
\centering
\includegraphics[width=.95\textwidth]{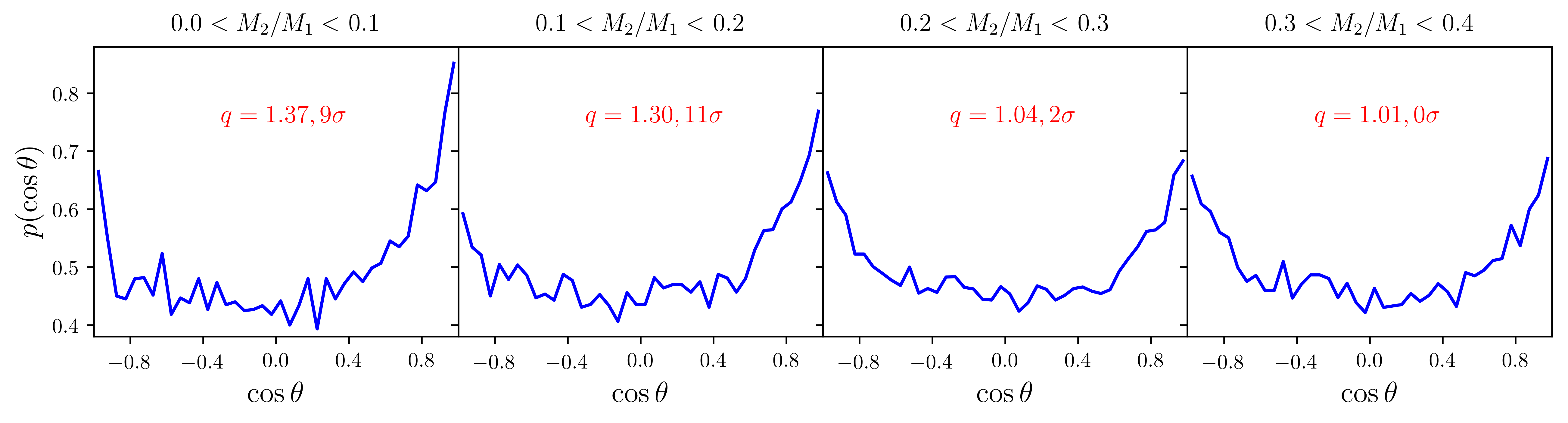} 
\caption{The lopsided signal is shown here for satellites of the less massive host partner, split according to host mass ratio (defined as $M_{2}/M_{1}$ where $M_{2}<M_{1}$). The lopsided signal is strongest for the most extreme mass ratios. Systems of mass ratios larger than $0.4$ are not shown here as they do not show substantial differences from those in the lower mass ratio range (between 0.2 and 0.4). \label{fig:roche}} 
\end{figure*}

\subsubsection{Host mass and mass ratio} \label{sec:massratio}
Here we examine how the mass ratio (defined as $M_{2}/M_{1}$ where $M_{2}<M_{1}$) of the pair affects the lopsided satellite distribution. In Fig.~\ref{fig:density_bias2} we show the distribution of satellite position angle around the less and more massive haloes independently. Fig.~\ref{fig:density_bias2} reveals that the more massive hosts show a weaker lopsided signal than the less massive partners. 

We now examine the effect of the mass ratio on the probability distribution of $\cos\theta$ for the less massive partner. The reader will recall from section \ref{sec: haloid} that the haloes in our sample may have masses anywhere in the range $10^{11} \sim 5\times10^{12}h^{-1}M_{\odot}$, i.e. may have a mass ratio from 0.02 to unity. In Figure \ref{fig:roche} our sample is divided according to the mass ratio of the pair, and the distribution of $\cos\theta$ for the satellites of the less massive partner (only) is shown. Figure \ref{fig:roche} shows that the lopsided satellite signal is strongest for the smallest mass ratios (for example when the more massive halo is 4 or 5 times larger than its smaller partner). 

We conclude this section by stating that the lopsided signal is greatest in the less massive halo pair, and for the most extreme pair halo mass ratios. However, as in the previous section, we again ask the question if the signal could be due to overlapping effects, especially suspect are the halo pairs with large mass ratios, since the more massive halo could influence its environment out to a few virial radii. Thus we construct artificial pairs that match the mass ratio bins shown in Fig.~\ref{fig:roche} and examine the strength of the lopsided effect due to overlap. In all cases the effect is indistinguishable from the distributions shown in Fig. \ref{fig:2Ddensity_all}b (not shown). Given that the distribution in the most extreme mass ratio is highly lopsided ($q=1.37$), it is impossible that such an overlap effect can cause the lopsidedness shown in Fig.~\ref{fig:roche}.

\section{The origin of the lopsided signal} \label{sec: highz}
In section \ref{sec: signal} and \ref{sec: quantsignal} we established the existence of the lopsided satellite distribution in pairs of haloes in a cosmological simulation at $z=0$. We now attempt to uncover the origin of these distributions by following each satellite's orbital trajectory back in time. To do so, we use the {\sc Rockstar} Mergetree software described in section \ref{sec: Rockstr}.

\subsection{The role of infall}
In order to examine the origin of the $z=0$ lopsided signal we wish to trace back in time the most massive progenitor of all pairs and their satellites. However, not all satellites can be traced  back through previous snapshots, since some of the smaller satellites will disappear as their mass fall below the halo finder's resolution. In order to keep the sample size at each redshift the same, we are thus forced to cull from our sample those satellites which during the back-tracking procedure fall below the resolution limit and disappear from the merger tree. We arbitrarily set the redshift to which we wish to track orbits to $z=2.5$ which maximises the sample size and the redshift. Using $z=2.5$, we are forced to cull $\sim35\%$ of the subhaloes that are found at $z=0$ as these fall below the resolution limit at some point between $z=0$ and 2.5.

\begin{figure}
\centering
\includegraphics[width=.4\textwidth]{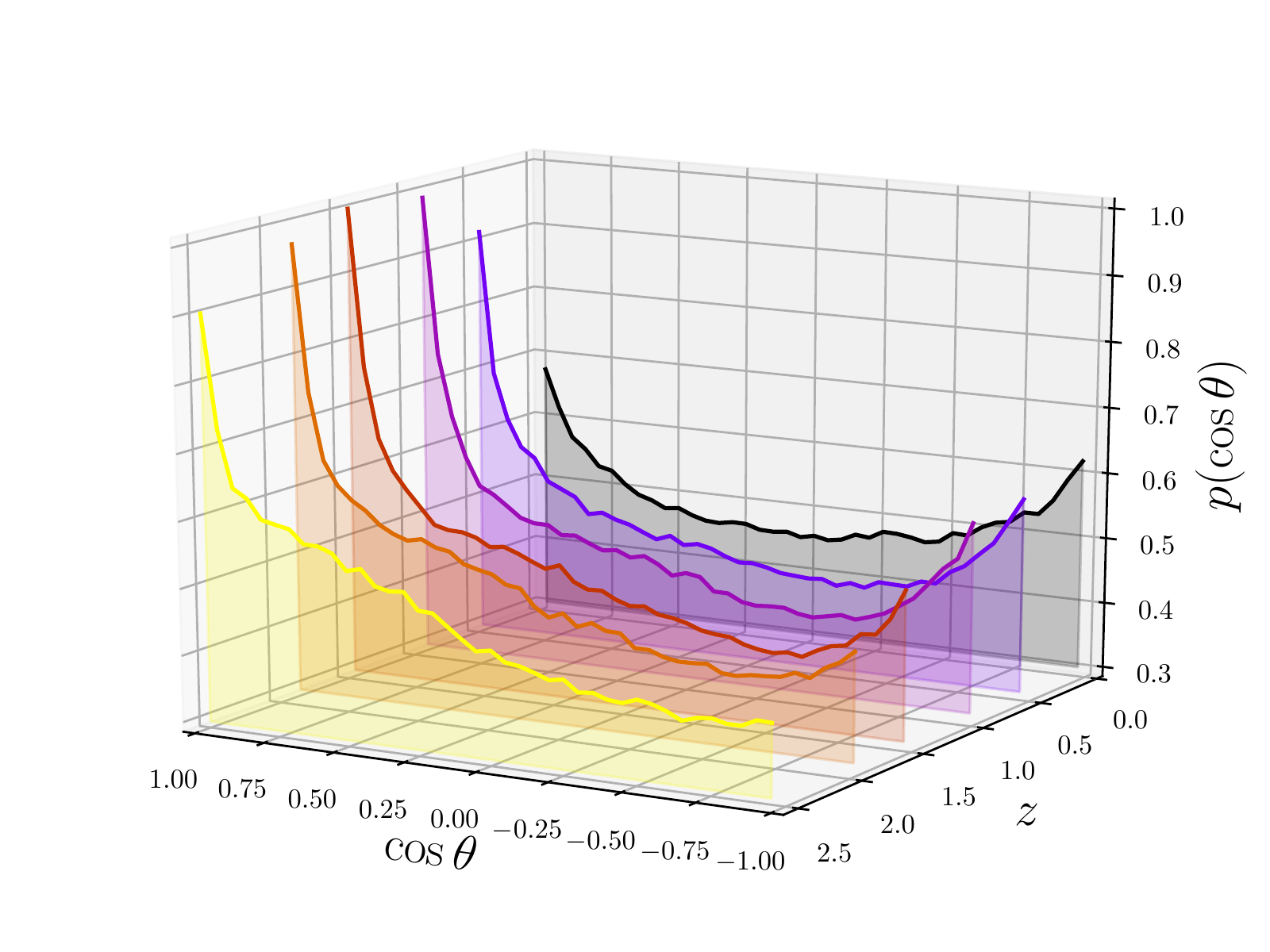}
\caption{The evolution of the cosine of the angle $\theta$ as described in Fig.~\ref{fig:bindraw}. The lopsided signal distribution is stronger in the past than it is today.  \label{fig:stackedA}}
\end{figure}

\begin{figure}
\centering
\includegraphics[width=.4\textwidth]{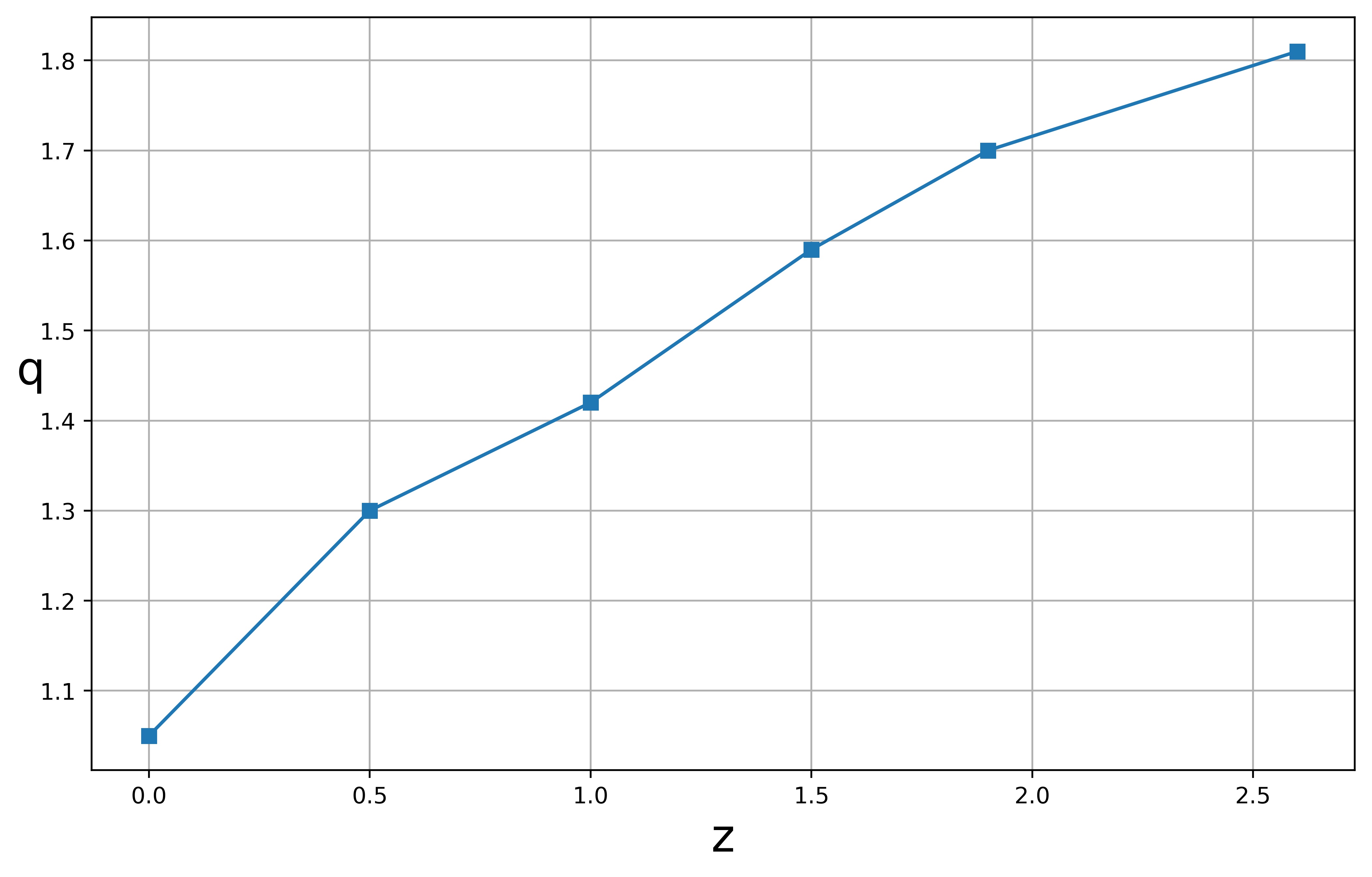}\\
\caption{The evolution of the lopsided signals is described by the lopsided metric $q$, shown here as a function of $z$ for all pairs considered in Fig.~\ref{fig:stackedA}.\label{fig:qvsz}}
\end{figure}

In Fig.~\ref{fig:stackedA} we present five probability distribution of this sample as it is traced back through cosmic time, at $z=0,0.5, 1, 1.5, 2$ and 2.5. The reader will first note that this culled sample still shows a lopsided signal at $z=0$. It is immediately apparent that the lopsided signal was significantly stronger in the past and that there is a considerable weakening as the system evolves towards $z=0$. Namely, satellites of $z=0$ pairs, are more likely to be accreted from (and therefore to have formed in) the regions between the two hosts. Satellites are less likely to be accreted from (or to have formed in) the regions diametrically on the other side, opposite the host's partner. The weakening of the lopsided signal as cosmic time advances may be quantified by examining  the lopsided metric $q$ as function $z$, shown in Fig.~\ref{fig:qvsz}; a monotonically decreasing trend is seen.

\begin{figure}
\centering
\includegraphics[width=.4\textwidth]{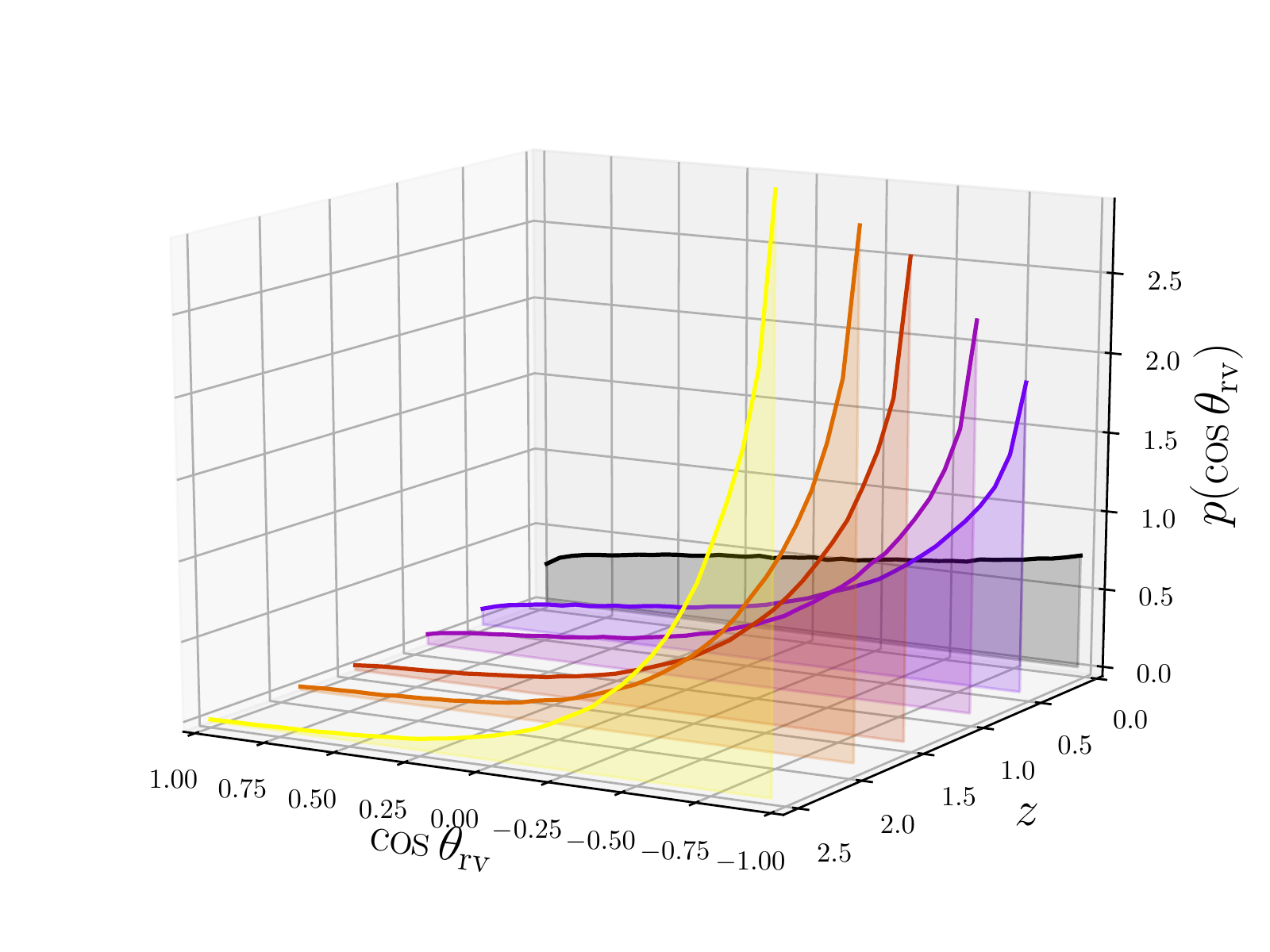}\\
\caption{Evolution of cosine velocity incident angle distribution of satellite haloes.  \label{fig:stackedB}}
\end{figure}

In Fig.~\ref{fig:stackedB} we examine $\cos\theta_{\rm rv}$, the angle formed between a satellite's position vector (with respect to the most massive progenitor of its $z=0$ host), and its (relative) velocity vector. Namely at each redshift under consideration, we examine how radial the satellite's velocity vector is with respect to its host. Note that a satellite with $\cos\theta_{\rm rv} = -1 $ is moving radially towards its host with 0 angular momentum; a satellite with $\cos\theta_{\rm rv} = 1$ is moving radially away from its host, also with zero angular momentum, while a satellite with $\cos\theta_{\rm rv} = 0$ is moving perpendicularly or tangentially to its position vector. A coherent picture emerges: at higher redshifts, the distributions of $\cos\theta_{\rm rv}$ is highly skewed, indicating a strong tendency for radial accretion. At low $z$, as satellites approach their host, the velocity vector no longer has such a strong tendency to be aligned with the position vector. This is likely due to a combination of the gravitational pull of the host's partner (as the pair separation shrinks) as well as other ``tidal'' or nonlinear virial effects due to the increase in number density of other satellite galaxies.


\begin{figure*}
\centering 
\includegraphics[width=.95\textwidth]{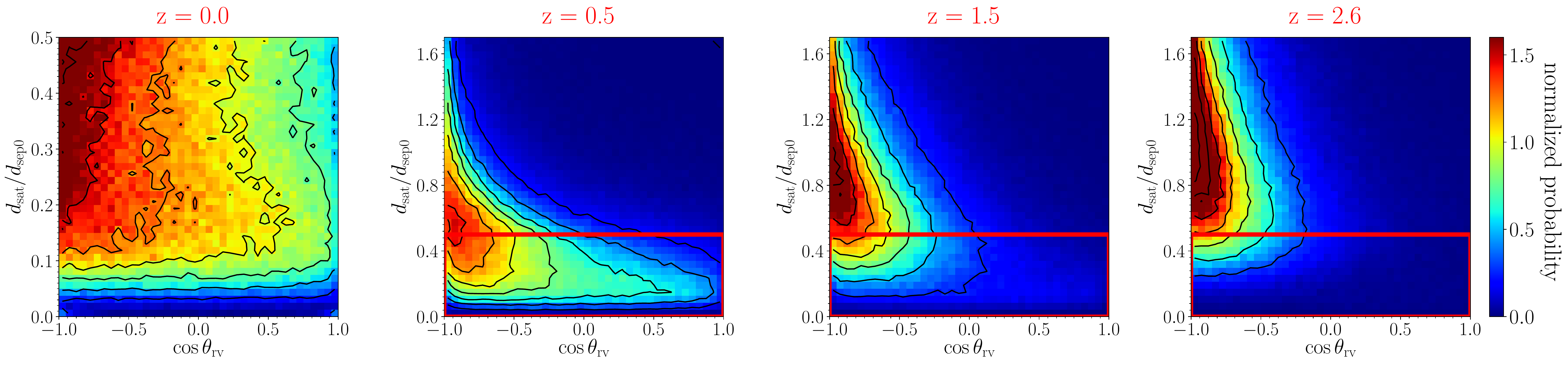}\\
\caption{Normalized 2D histogram for the distribution of the radial infalling angle $ \theta_{\text{rv}}$ as a function of satellite distance from host (in units of $d_{\rm sep0}$, the $z=0$ pair separation) at various redshift. Colours are used to visualise which regions of these planes are more populated by data points. Note that the sample can not extend beyond $0.5 d_{\rm sep0}$ at $z=0$, by construction. For higher redshift the red line is used to mark the region that is searched at redshift 0. \label{fig:densityplot_merger}} 
\end{figure*}

In Fig.~\ref{fig:densityplot_merger} we show how a satellite's velocity alignment angle ($\cos\theta_{\rm rv}$, see Fig.~\ref{fig:bindraw}) depends on its distance (measured in units of the pair separation at $z=0$,  i.e. $d_{\rm sep0}$) for four different redshifts, $z=0, 0.5, 1.5$ and 2.5. Fig.~\ref{fig:densityplot_merger} shows that when satellites are very distant, most  come in on ``radial-like'' orbits i.e. with $\cos\theta_{\rm rv} \approx -1 $. As $z$ goes to 0, $\cos\theta_{\rm rv}$ begins to fill the interval [-1,1], which might be an indication that many of the satellites gain angular momentum. At $z=0.5$, where the region of satellite moving away from their host (namely with $\cos\theta_{\rm rv} > 0$) begins to be filled, we see that it is almost entirely populated by satellites located at distances of $< 0.5 d_{\rm sep0}$. Perhaps most pertinent to our study, we find that by $z=0$ up to $\sim$40\% of the satellites are actually moving away from their hosts (i.e. $\cos\theta_{\rm rv} > 0$), after they probably have flown by their hosts at some point in the past. 

Since Fig.~\ref{fig:densityplot_merger} indicates that when, for the most part, the lopsided signal is strong, the angular momentum might be low and satellites are travelling on infalling orbits (and vice versa), we hypothesize that an encounter between the accreted satellite and its host may be responsible for both weakening the lopsided effect and generating orbital angular momentum of the satellite. This is a well established aspect of the virialization process which accompanies halo relaxation, namely the stabilization of angular momentum from infalling material (\citealt{Stewart13}).

\subsection{Flybys and their impact on the lopsided signal} \label{sec:flyby}

In order to test the hypothesis that accreted satellites that undergo ``flyby'' event weaken the lopsided effect of satellite distribution, we split the satellite sample up into those satellites that have ``flown by'' their host and those that haven't. This is done by following back in time the trajectory of each satellite, and by  examining the distance between the satellite and its host. If this distance experiences a local minimum (that is not at $z=0$) then the satellite has flown by its host. 

\begin{figure}
\centering
\includegraphics[width=.5\textwidth]{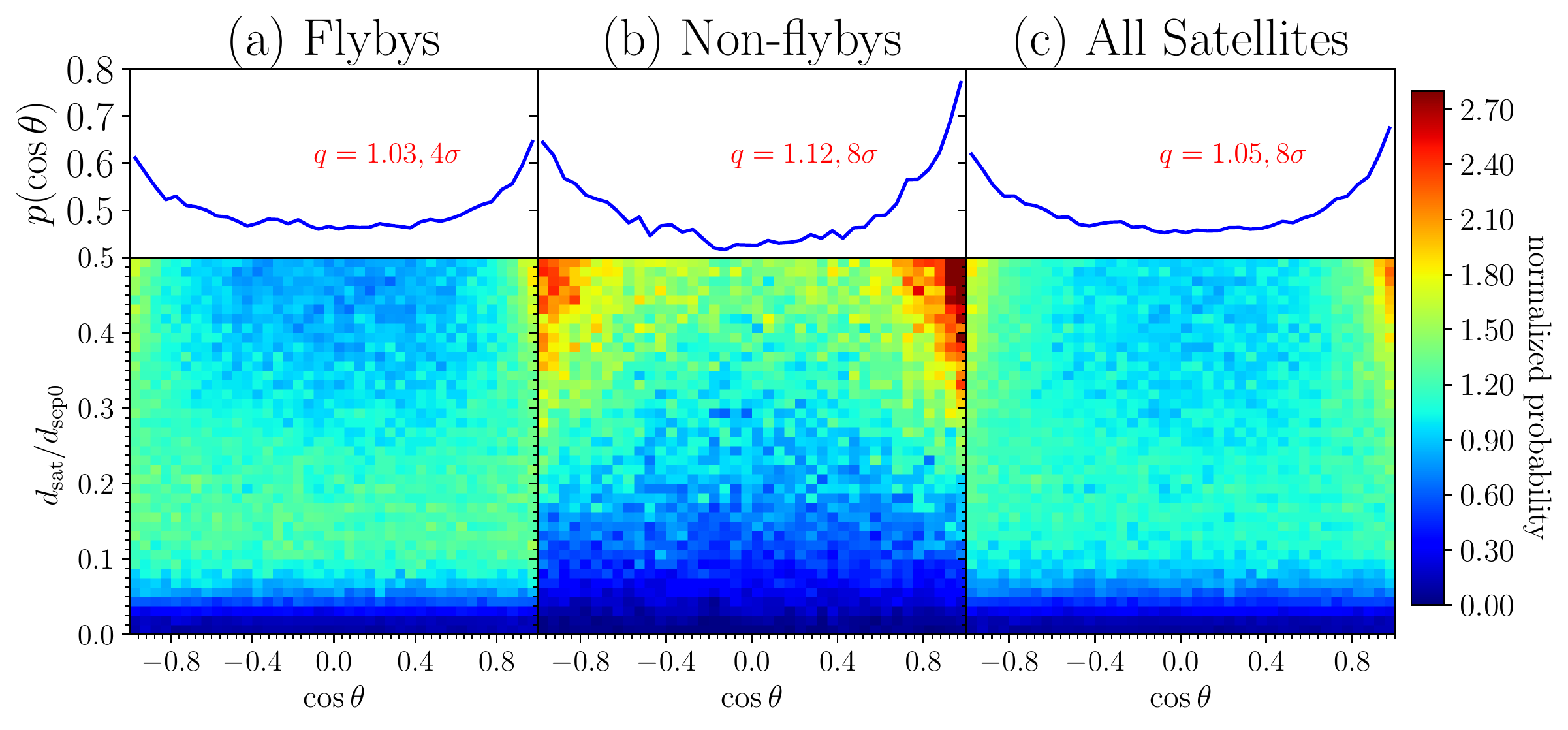}\\
\caption{Flyby signal (a) weakens the total lopsidedness of the signal (c) at z = 0. \label{fig:flyby_signal_0} }
\end{figure}

In Fig.~\ref{fig:flyby_signal_0} we present the $z=0$ probability distribution of $\cos\theta$ for those satellites labelled as flybys (a) and for those that have not experienced an encounter (b). For completeness we show the same plot for the full $z=0$ sample\footnote{The reader will note that this is identical to the black curve in Fig.~\ref{fig:stackedA}. It is not the same as the probability distribution shown in Fig.~\ref{fig:binwsigma} or \ref{fig:2Ddensity_all}a since it includes only those satellites that can be traced back in time to $z=2.5$.}. Fig.~\ref{fig:flyby_signal_0} strikingly demonstrates  the difference in these two samples. Those satellites that have incurred a flyby at one point in the past, are both more centrally concentrated and therefore (see Fig.~\ref{fig:linedensity}) more uniformly distributed, and less lopsided. Those satellites that have not experienced a close flyby are both more aligned with $\cos\theta\approx\pm1$ and more lopsided. The lopsided signal appears to strongly be driven by satellites that are infalling and who have not been gravitationally slung about their host. 

Note that the flybys have not fully erased the lopsided signal, they have merely reduced the total lopsided signal to a very weak degree ($q=1.051$), so weak in fact, that it is consistent with the effect of overlapping haloes. Nevertheless, as we have seen so far, the overlapping effect alone cannot explain the lopsidedness in the signal given that it weakens with decreasing redshift.  

\section{Discussion and Conclusions}
The conclusion we draw regarding the origin of the lopsided signal is the following. The satellites that cause the signal tend to be formed in between the pair, in the filament connecting the pair. These satellites are then attracted to each member of the pair due to gravity. Although those satellites that are accreted and ``swung'' to the other side of the attracting pair member may end up preferentially closer to the line connecting the binary, they are outnumbered by the satellites still being accreted from the region between the pair. The filamentary formation and accretion scenario may be related to the 3-point correlation function \citep{Peebles80} which has been measured in the SDSS by \cite{McBride11} but which shows only a very weak signal below $\sim 6 h^{-1}$ Mpc, \citep[see also][]{Ma00,Jing04}, due mostly to non-linear effects (such as flybys). 

We examined how simply overlapping two satellite distributions would add to the observed signal. By constructing artificial pairs we found that indeed at z=0 overlapping satellite distributions have  lopsidedness that is on the same order of magnitude as that of the observed haloes. We conclude that, the effects described above, namely that the signal is weakened by satellites that have had a pericentric passage and that the lopsidedness is stronger at higher z, rule out overlapping satellite distributions as being primarily responsible. In summary: the lopsided signal originates at high redshift and is weakened as the halo pair distance shrinks and satellites are accreted and have pericentric passages. Such dynamical effects weaken the signal to the degree that at z=0 it is consistent with artificially created overlapping satellite distributions but not due to it.

As mentioned in Sec.~\ref{sec: intro}, \cite{Libeskind16} examined how satellites were spatially distributed in spectroscopically identified SDSS pairs. \cite{Pawlowski17} went on to examine whether such satellite distributions could be found in $\Lambda$CDM cosmological simulations, coming to the conclusion that indeed such a lopsided distribution was consistent with what the cosmological paradigm predicted. In order to follow up on the question of ``why do satellites tend to be in the in-between regions of galaxy pairs?'', in this work we examined how pairs of galaxies identified in cosmological simulations at $z=0$ accrete satellites. By employing a large, well resolved, cosmological simulation we are able to numerically follow the formation of such pairs and their satellites. We trace back each host progenitor and each satellite to $z=2.5$. At each snapshot we construct the triangle composed of the (progenitors of the) two members of the binary and the (progenitor of the) satellite and examine the evolution of the angle satellite -- host -- host partner. We find that the signal is strongly driven by satellites that are on their first approach to the binary pair (by ``first approach'' we mean satellite that have never been closer to their host). Satellites that were at some point in the past closer to their hosts (a category we call ``flyby'' satellites) exhibit a much weaker  lopsided distribution. Lopsided distributions of satellite galaxies among pairs of galaxies are thus characterised by dynamically young accretion events that have either recently come to the pair, or are still in the process of being accreted. This is similar to the picture described by \cite{2011MNRAS.412...49W} where accretion is more radial at higher $z$. We have also found that the lopsided distribution is more often found in the lower mass member of the pair, where lower mass implies lower degree of virialization and hence a stronger lopsided signal, especially when its partner is much larger. A higher mass member of the pair has in general more satellites which are accreted, and hence corresponds to a denser environment. 

The pairs examined in this paper (or indeed in other works that focus on the lopsided signal) are not identical to the Local Group or the Centaurus group, although such a lopsided distribution is also seen in the distribution of satellites along Andromeda where 19-23 out of 27 satellites are on the near side of M31 (\citealt{2013Natur.493...62I}) and (albeit to a lesser extent) around the Milky Way where the census is poorer due to obscuration from the galactic disk and inhomogeneous coverage from sky surveys such as the SDSS \citep[]{2012AJ....144....4M}, as well as in the Centaurus A - M83 pair (\citealt{Mueller15}). That being said, the selection criteria employed here is loosely motivated such that the Local Group would also be identified according to the definitions we use. Our criteria are purposefully kept more relaxed such that we may get a more general picture of these systems.

Our work confirms one of the assertions of \cite{Pawlowski17} that lopsided distributions are not in conflict with $\Lambda$CDM models of structure formation, at least when comparing surveys like the SDSS to the virialized haloes that form in dark matter only $N$-body simulations. However, our discovery that the origin of these systems is primarily newly accreted satellites is somewhat surprising, since its well known that satellites may orbit their hosts for many gigayears before being accreted \citep[for example:][among others]{1998MNRAS.299..728T,2009ApJ...693..830Y,2019MNRAS.tmp..397B}. 

We leave the reader with the following notion regarding satellites of ``Local Group like'' pairs: if the existence of lopsided distributions in surveys and simulations is driven by satellites on first approach, we infer that, either these are a different population and unlike satellites found within the virial radius of Milky Way like haloes that may survive for many giga years, or the lopsided signal in Local Group is also driven by such mostly accreted satellites. Indeed, recent work by \cite{Hammer18} suggests that this may be the case and that the satellites of the Milky Way may indeed be on their first approach.

\section*{Acknowledgments}
NIL acknowledges financial support of the Project IDEXLYON at the University of Lyon under the Investments for the Future Program (ANR-16-IDEX-0005). NIL, PW and QG, acknowledge support from the joint Sino-German DFG research Project ``The Cosmic Web and its impact on galaxy formation and alignment'' (DFG-LI 2015/5-1). ET was supported by ETAg grants (IUT40-2, IUT26-2) and by EU through the ERDF CoE grants TK133 and MOBTP86. GY acknowledges support  by  Ministerio de Economia y Competitividad and the Fondo Europeo de Desarrollo Regional (MINECO/FEDER, UE) in Spain through grants AYA2015-63810-P and PGC2018-094975-B-C21. JS acknowledges support from the ``Centre National d'\'{e}tudes spatiales (CNES)'' postdoctoral fellowship program. The ESMD simulation has been performed at LRZ Munich within the project pr74no. 


\bibliographystyle{mnras}
\bibliography{biblio} 
%


\bsp	
\label{lastpage}
\end{document}